\documentclass[twocolumn,showpacs,preprintnumbers,amsmath,amssymb]{revtex4}
\usepackage{amssymb}
\usepackage{graphicx}
\usepackage{latexsym}

\begin{document}
\title{Experimental investigation of double layers in expanding plasmas}

\author{ N. Plihon}
\email{plihon@lptp.polytechnique.fr} \affiliation{Laboratoire de
Physique et Technologie des Plasmas, Ecole Polytechnique, $91128$,
Palaiseau Cedex, France}
\author{P. Chabert}
\affiliation{Laboratoire de Physique et Technologie des Plasmas,
Ecole Polytechnique, $91128$, Palaiseau Cedex, France}
\author{C. S. Corr}
\affiliation{Laboratoire de Physique et Technologie des Plasmas,
Ecole Polytechnique, $91128$, Palaiseau Cedex, France}

\begin{abstract}

Double layers (DLs) have been observed in a plasma reactor composed of a source chamber attached to a larger expanding chamber. Positive ion beams generated across the DL were characterized in the low plasma potential region using retarding field energy analyzers. In electropositive gases, DLs were formed at very low pressures (between 0.1 and 1 mTorr) with the plasma expansion forced by a  strongly diverging magnetic field. The DL remains static, robust to changes in boundary conditions, and its position is related to the magnetic field lines. The voltage drop across the DL increases with decreasing pressure, i.e. with increasing electron temperature (around 20 V at 0.17 mTorr).
DLs were also observed in electronegative gases without a magnetic field over a greater range of pressure (0.5 to 10 mTorr). The actual profile of the electronegative DL is very sensitive to external parameters and intrusive elements, and they propagate at high negative ion fraction. Electrostatic probes measurements and laser induced photodetachment show discontinuities in all plasma parameters (electron density, electron temperature, negative ion fraction) at the DL position. The voltage drop across the electronegative DL is about 8 V, is independent of the gas pressure and therefore of the electron temperature. 
\end{abstract}

\maketitle
\date{\today}
\newpage

\section{Introduction}

An electric double-layer (DL) is a region
within a plasma in which localized charged regions can sustain a potential difference and
cause electron and ion acceleration~\cite{Raadu89}. A DL can be thought of as an internal sheath separating plasmas at different potentials. Following  general terminology, in this paper the high plasma potential region is referred to as the "upstream" region and the low plasma potential region across the DL as the "downstream" region. DLs occur naturally in a variety of space plasma environments and are of great interest in astrophysics. In the laboratory, they have been experimentally studied for decades. In most of the experiments, the DLs were created in double plasma devices where two plasmas were externally biased at independent potentials. Due to the electrical bias, a current is flowing through these current-driven DLs. Since the work of Charles and Boswell~\cite{Charl03a}, there has been considerable interest in static current free double layers that are self-consistently created  in low-pressure expanding plasmas. These DLs have applications in space propulsion~\cite{Gesto06,Fruch05b} and in modeling of astrophysical processes~\cite{Boswe06}. In these experiments, a plasma expansion was forced by a strong divergence of a high magnitude magnetic field at very low pressure. Detailed investigations of these current-free DLs have been performed using electrical measurements~\cite{Charl03a,Charl04c,Charl04b,Charl04,Charl05,Charl05a,Suthe05}, laser induced fluorescence~\cite{Cohen03,Sun04,Sun05b,Biloi05, Keese05} and numerical simulations~\cite{Meige05,Meige05a}. An upstream ionization instability has recently been observed in one of the experimental systems~\cite{Aanes06}. Despite the strong interest shown in this area, no general theory has been developed. A recent model of DL formation has been constructed by Lieberman and co-workers and compared with experiment~\cite{Liebe06}. Chen showed that, based on a one dimensional model, the physical process is the same as for single layer formation, however no general explanation of how the single layer develops into a DL is given~\cite{Chen06}. 

In this article, we present an experimental study of the formation of a DL in a low pressure radio frequency driven electropositive plasma where its expansion is forced by a strongly diverging magnetic field. The conditions for DL formation and the evolution of the plasma parameters are analyzed using electrical measurements and compared to other investigations. Throughout the text 
these DLs are termed "electropositive DLs".

In a recent letter, we showed that a DL could also be created in an expanding plasma due to the presence of negative ions, without a static magnetic field, and at higher pressures~\cite{Pliho05}. Electronegative plasmas are widely studied not only because of their relevance for processes in the microelectronic industry, but also because their fundamental properties are significantly different from 
electropositive discharge plasmas~\cite[chap. 8]{Liebe05}. Some of the most important and interesting problems are related to the existence of various kinds of non-linear potential structures and instabilities. Nonlinear potential structures occur in plasma stratification between an electropositive outer shell (no negative ions in the plasma) and an electronegative core (negative ion containing plasma). This feature was theoretically studied~\cite{Ferre88,Frank92,Licht94,Sheri99a,Lampe04} and experimentally observed~\cite{Vende95,Berez00}. Furthermore, it has been shown that inductive discharges are subject to relaxation oscillations near the capacitive to inductive mode transition (at intermediate powers) in the presence of negative ions~\cite{Chabe01,Chabe03,Corr03,Corr05}. 
At sufficiently high power (in the inductive mode), the plasma is not subject to these relaxation oscillations. However, in an expanding plasma, a periodic propagating phenomena had been identified by Tuszewski and co-workers~\cite{Tusze03,Tusze03b}. Plihon and co-workers~\cite{Pliho05a} showed that this periodic propagating structure is a double layer for their experimental conditions. We remind here 
 the basic features of this phenomenon in Ar/SF$_6$ gas mixtures. When the SF$_6$ concentration is above 13\%, the discharge is non stationary, with plasma parameters fluctuating due to the periodic formation and propagation of a double layer occuring at the interface between the source and the diffusion chamber. For lower SF$_6$ concentrations (namely between 8 and 13\% at 1.5 mTorr), the double layer appears to be static, and eventually disappears for SF$_6$ concentrations below 8\%. The present article details an extensive experimental study of the static electronegative DL, in the same experimental device as the one used for the study of electropositive, magnetized DLs. Ar/SF$_6$ and Ar/O$_2$ gas mixtures are investigated. Although the physical processes leading to the formation of the electronegative DLs are different from those involved in the formation of the electropositive DL, we provide a comparison for the two cases.

This paper is divided into five parts organized as follows. Part II presents the experimental setup and diagnostics used to determine the plasma parameters. Part III presents a parametric study of the electropositive DL in a low pressure (below 1 mTorr), strongly magnetized plasma, and emphasizes similarities with previously published work. Part IV gives a detailed experimental investigation of the electronegative DL in Ar/SF$_6$ and Ar/O$_2$ gas mixtures. Some conclusions are given in part V.

\section{Experimental setup and diagnostics\label{sec.exp}}

\subsection{The plasma source}

\begin{figure}
\includegraphics[width = 0.8\linewidth]{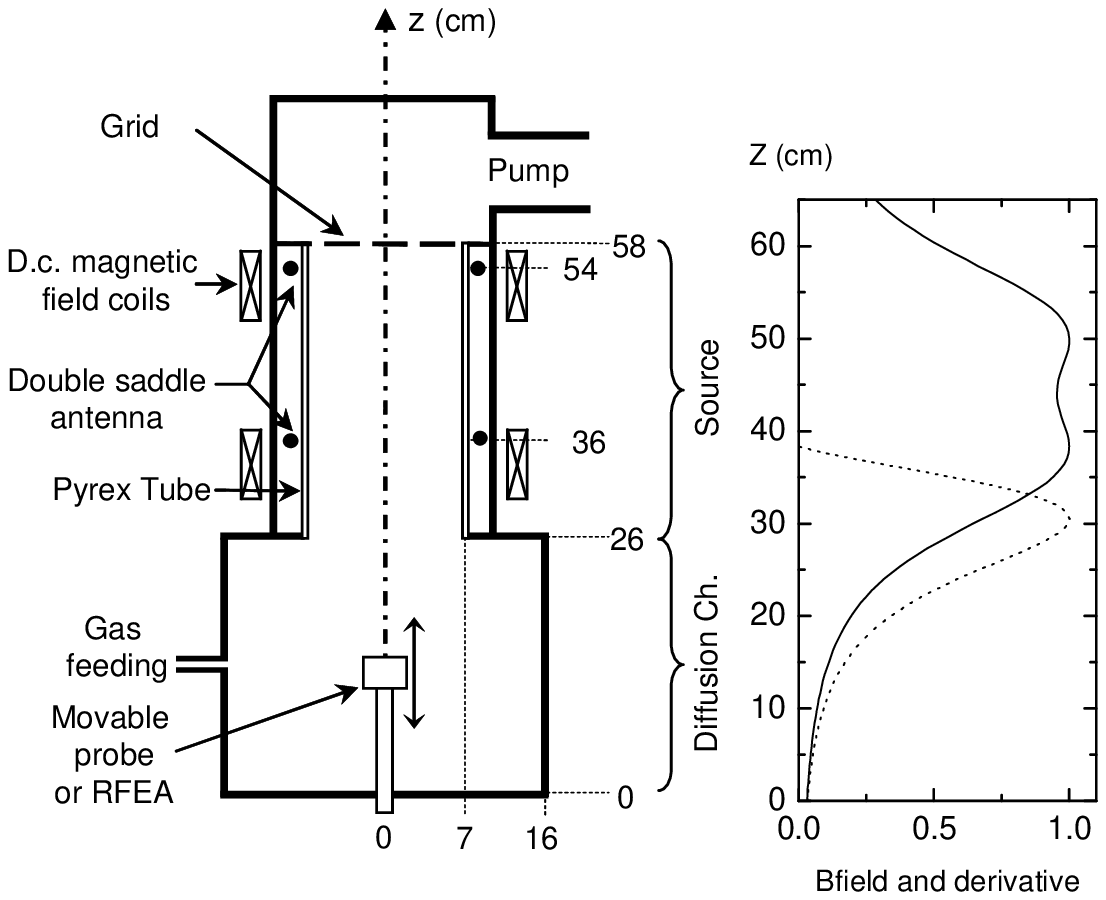}
\caption{\label{Schema}Schematic of the experimental device and normalized axial magnetic field magnitude (solid line) and derivative (dots) used for investigation of the magnetized electropositive DL in section~\ref{sec.HDLT}.}
\end{figure}

The plasma source is shown schematically in figure~\ref{Schema}. It
consists of a source chamber on top of a 32 cm diameter
diffusion chamber. 
The source is a 15 cm diameter, 30 cm long and 0.9 cm thick pyrex
cylinder surrounded by a double saddle field type helicon antenna
\cite{Boswe70}. The fan-cooled antenna is powered through a
close-coupled L-type matching network by an rf power supply
operating at 13.56 MHz and capable of delivering up to 2 kW
forward power. The time-averaged input power was recorded as the
difference between the time-averaged forward and reflected powers.
The pyrex cylinder is housed in a 20 cm
diameter and 30 cm long aluminum cylinder. A grid attached to the other end of
the source tube confines the plasma and isolates it from a
turbo-molecular pump that routinely maintains base pressures of
$10^{-6}$ mbar. The boundary conditions terminating the plasma can be modified, as specified in section~\ref{sec.boundary}. The discharge was operated in pure Ar, Ar/$\text{SF}_6$ and Ar/$\text{O}_2$
gas mixtures. The partial gas pressures of Ar and the molecular gases were
determined by controlling the flows and the total pressure was measured using a Baratron pressure gauge.

A static magnetic field can be generated by a d.c. current in  two coils surroundinfg the source tube. The coils are designed such that the magnetic field is axial and uniform in the source and strongly divergent at the plasma source - diffusion chamber junction. Figure~\ref{Schema} shows the normalized axial magnetic field magnitude (solid line) and the derivative (dots). Results presented in section~\ref{sec.HDLT} were obtained with a d.c. magnetic field in a pure argon plasma, while in section~\ref{sec.ENDL} the reactor was operated without a d.c. magnetic field. The magnitude of the magnetic field mentioned in the text is the maximum field obtained along the source axis.

\subsection{Langmuir probe operation}

The measurements reported here were performed along the revolution
axis ($z$ axis) of the discharge. All probes can also be inserted in the mid-plane of the diffusion chamber and radial dependence of the plasma parameters can be measured. The plasma parameters were determined using a passively compensated
Langmuir probe \cite{Canti77}, of 0.25 mm diameter and 6 mm long
platinum wire tip. The plasma potential, 
electron density and electron temperature were deduced from
the I(V) characteristics of the cylindrical probe using a
Smartsoft data acquisition system \cite{Hopki86}. The plasma parameters were also determined from usual I-V curve processing~\cite{Magnu02}: the plasma potential from the zero of the second derivative of I-V traces, the plasma density from the electron current at the plasma potential, and the electron temperature from a linear  fit of the logarithm of the electron current (assuming a maxwellian electron distribution function).
The electron energy probability function was determined from the second derivative of the I-V curves~\cite[p. 189]{Liebe05}
\begin{equation}
\text{eepf}(V) = \frac{2\sqrt{2 e m_e }}{e^2A_p}\frac{d^2I_e}{dV^2}
\end{equation}
where $e$ is the electric charge, $m_e$ the electron mass, $A_p$ the probe collection area and $I_e$ the electron current. The electron current was taken to be the difference between total collected current and positive ion saturation current, which is assumed to scale as the square root of the biasing voltage relative to the plasma potential~\cite{Mott-26}. According to~\cite{Magnu02} we computed the second
derivative of the I(V) curves using a Savitzky-Golay (S.G.)
filter~\cite{Savit64} of order 3, over 7 points. Since the filter works only for equally
spaced voltage points, we computed the derivatives with respect to
the supply voltage and used $\frac{d^2I}{dV_{\rm probe}^2} =
\frac{d^2I}{dV_{\rm supply}^2}\left(1 - R\frac{dI}{dV_{\rm
supply}}\right)^{-3}$~\cite{Sudit93} to calculate the
derivatives with respect to the probe voltage. The electron density was also obtained from the integration of the eepf 
\begin{equation}
n_{e\ \text{eepf}} = \int_0^{+\infty}\text{eepf}(V) V^{1/2} dV
\end{equation}

 An effective electron temperature was determined from the mean electron energy calculated from the eepf
\begin{equation}
T_\text{eff} = \frac{2}{3n_{e\ \text{eepf}}}\int_0^{+\infty}\text{eepf}(V) V^{3/2} dV
\end{equation}

When operating with the molecular gases, the plasma can contain a large number of negative ions. It is possible to determine the negative ion fraction $\alpha$, defined as the negative ion density over the electron density ratio, using a two probe technique~\cite{Chabe99}. This technique is based on the comparison of the Langmuir probe current at the plasma potential (the electron thermal current) with the positive ion saturation current from a planar probe, obtained in both an electropositive and an electronegative plasma. The calculation relies on a sheath model developed in~\cite{Sheri99a} and requires an estimate of the negative ion temperature.

\subsection{Retarding field energy analyzer}

In order to investigate the positive ion distribution function, we have developed a retarding field energy analyzer (RFEA) consisting of four grids and a collector
plate~\cite{Ingra88,Boehm93,Perre05}. The four grids are made of
nickel wires of 11 microns in diameter spaced by 40 microns; each grid had a 60 percent
transparency. The entrance grid was grounded, the second grid was biased at -50 V to repel
the electrons, the third grid was used to select the ion energy by scanning the voltage from -10
to 60 V, and a fourth grid was inserted before the collector in order to minimize the effect of
secondary electrons. The grids were spaced by 0.25 mm so that the total system length
was about one millimeter. When assembled, the analyzer was 35 mm long by 50
mm in diameter and the plasma particles enter the analyzer through a 2 mm hole in a 0.3
mm thick stainless steel orifice plate. The plate is in electrical contact with the analyzer housing, which
is connected to the grounded diffusion chamber of the reactor. The RFEA is supported by a 3 cm diameter aluminum tube that is differentially pumped and can be moved along the vertical axis.
The ion current to the collector $I_c$ was recorded as a function of the dc voltage applied to the discriminator $V_d$ and is given by:
\begin{equation}\label{RFEAcurrent}
\begin{aligned}
I_c(v_0) = e A_R G^4 \int_{v_0}^{\infty}v f(v) dv \\
\text{with}\qquad v_0 = \sqrt{\frac{2e V_d}{M_+}}
\end{aligned}
\end{equation}
where $A_R$ {is} the collecting surface, $G$ the grid transparency, $v_0$ the speed
of the ion accelerated by the discrimator grid at potential $V_d$, $f(v)$ the ion distribution function and $M_+$ the positive ion mass. The power 4 of $G$ accounts for the transmission across the four grids. The ion
energy distribution function (IEDF) was computed as \begin{math} -\frac{d I_c}{d V_d}\end{math}. When operating in a low pressure plasma, the ion distribution function is centered around the plasma potential with a dispersion due to the ion temperature or the RFEA resolution (in our case the RFEA resolution is such that the ion temperature cannot be measured). 
For measurements downstream of a DL, a double peaked IEDF is
expected, with one peak centered at the plasma potential, and the other corresponding to
an ion beam;  the positive ion saturation current is thus the sum of (i) an ion beam current from ions created upstream and accelerated within the DL, and (ii) a local ion current from ions created in the downstream plasma. 
In order to probe the temporal development of the IEDF when pulsing the plasma, we developed a time resolved data acquisition system. The discriminator grid biasing voltage was kept constant during 256 plasma pulses and increased step by step afterward. The collected current was recorded using a large memory digital oscilloscope, averaging over 256 pulses to reduce noise (the temporal behavior was checked to be reproducible between each pulse). Voltage-current characteristics were then reconstructed and the time resolved IEDFs were computed.

\subsection{Probe based laser induced photodetachment}

When operating the discharge in electronegative gas mixtures, the negative ion fraction $\alpha$  was measured by probe based laser induced photodetachment (LIP)~\cite{Bacal00}.
The experimental arrangement consisted of a pulsed cylindrical laser beam aligned to be collinear with an uncompensated probe positively biased. The negative ion fraction is determined from measurements of the electron saturation current in the absence of the laser pulse ($I_{e\ dc}$, since the laser duty cycle is less than 10$^{-7}$) and the increase in the current ($\Delta I_e$) immediately after the pulse, using the relationship
\begin{equation}\label{eq.alpha} \frac{\Delta I_e}{I_{e\ dc}} = \frac{n_-}{n_e} = \alpha \end{equation}
where $n_-$ and $n_e$ are the  negative ion an electron
densities, respectively. The necessary conditions~\cite{Bacal00} for the laser beam diameter (set to be 6 mm), the laser beam power and probe voltage (set to be + 35 V) are set such that the negative ion fractions were measured correctly.
The laser photon energy required to detach an electron depend on the electron affinities of the molecular gas used (SF$_6$ or O$_2$). 

When operating in Ar/O$_2$ mixtures, O$^-$, O$_2^-$ and O$_3^-$ can be produced in high density low pressure plasmas, with electron affinities of 1.46 eV, 0.44 eV and 2.1 eV,
respectively~\cite{Neuma85,Drzai84}. Thus the photon energy from a frequency doubled Nd:YaG laser (532 nm - 2.33 eV) is sufficient to photodetach all negative ions.

When operating in Ar/SF$_6$ mixtures, negative ions are produced from collisions involving SF$_6$ fragments, and  F$^-$, SF$_3^-$, SF$_5^-$ and SF$_6^-$ are the dominant negative ion species~\cite{St-On00,Nagas95,Sasak97}. The electron affinity for these species are 3.4 eV~\cite{Blond89}, 2.88 eV ~\cite{Hotop85}, 3.79 eV~\cite{Chen88} and 3.16 eV~\cite{Chris00} respectively. However the electron affinity of SF$_5$ is still debated and might be as much as 4.2 eV~\cite{King96}. The use of a frequency-tripled Nd:YaG (355 nm) provides photons with sufficient energy to detach most of the negative ions (3.5 eV). It is possible that SF$_5^-$ is not photodetached and this will cause discrepencies with negative ion fraction determined by the two probe technique.

The laser power dependence of the LIP signal is given by
$$ \frac{\Delta I_e}{I_{e\ dc}}= 1 - \exp{\left(-\frac{\sigma P \Delta t}{h \nu A_L}\right)}$$
where $h \nu$ is the photon energy, $P$ is the laser beam power, $\Delta t$ the laser beam pulse length, $A_L$ is the laser beam section, and $\sigma$ the LIP cross section for photodetachment of a negative ion at a particular laser energy ($\sigma$  is zero for energies $h \nu$
lower than the electron affinity). From published cross sections, it is possible to identify the dominant negative ion in the discharge by fitting the beam power dependence of the LIP signal to theoretical dependence.

\section{Double layers in low pressure magnetized electropositive plasmas\label{sec.HDLT}}

In this section results obtained in an Ar plasma, at very low pressure and with d.c. current flowing in the source coils are presented. 
 The measurements  confirm the existence of DLs in these types of discharges and complements previous work.

\subsection{Existence of a double layer and spatial dependencies}

\begin{figure}
\includegraphics[width = 0.8\linewidth]{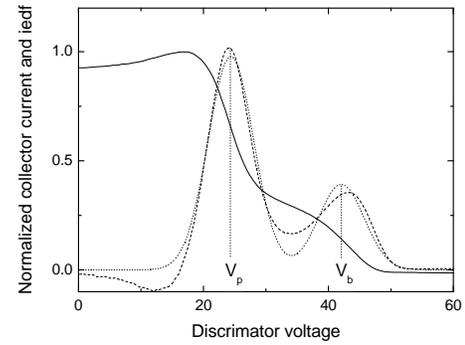}
\caption{\label{IEDFDL}Normalized RFEA IV curve (solid line), experimental IEDF (dashed line) and gaussian fit of the experimental IEDF (dotted line) showing evidence of an electropositive DL at 0.1 mTorr, 90 G and 250 W.}
\end{figure}

The RFEA was inserted into the diffusion chamber, facing the source. 
A typical current-voltage characteristic obtained from the RFEA is given in figure~\ref{IEDFDL} (solid line) when operating at sufficiently low pressure (below 1 mTorr) and moderate magnetic field (above 60 G). In order to provide a systematic analysis, the experimental IEDF (dashed line in figure~\ref{IEDFDL}) is fitted as the sum of two independent gaussian functions according to a least square procedure (dotted line). 

The gaussian functions are centered around $V_p$ and $V_b$ and their integrals are referred to as $I_p$ and $I_b$. The determination of the plasma parameters is processed as follows: (i) the plasma potential is the position of the first maximum $V_p$, (ii) the local ion flux ($\Gamma_p$) is the total flux ($\Gamma_+$) times the ratio $I_p/(I_p+I_b)$, (iii) the beam potential is the position of the second maximum $V_b$, (iv) the ion beam flux ($\Gamma_b$) is the total flux ($\Gamma_+$) times the ratio $I_b/(I_p+I_b)$. The total flux $\Gamma_+$ is deduced from the collected current when the discriminator is set to 0 V, corrected from the grid transparency according to equation~\ref{RFEAcurrent}. The total flux measured by the RFEA was verified against planar probe flux measurements. The potential drop across the DL is computed as  $\Phi_{DL} = V_b-V_p$. The first peak of the IEDF corresponds to ions that are accelerated at the Bohm velocity $v_{\rm Bohm}$ when entering the sheath in front of the RFEA; the local plasma density is computed as 
\begin{equation}
n_p = \frac{\Gamma_p}{ e G^4v_{\rm Bohm}}
\end{equation}
The second peak of the IEDF is related to ions created upstream of the DL that are accelerated within the DL. The beam velocity is computed as
\begin{equation}\label{eq.beamspeed}
v_b = \sqrt{\frac{2q(V_b - V_p)}{M_+}}
\end{equation}
and the beam density as
\begin{equation}
n_b = \frac{\Gamma_b}{ e G^4v_b}
\end{equation}

\begin{figure}
\includegraphics[width = 0.7\linewidth]{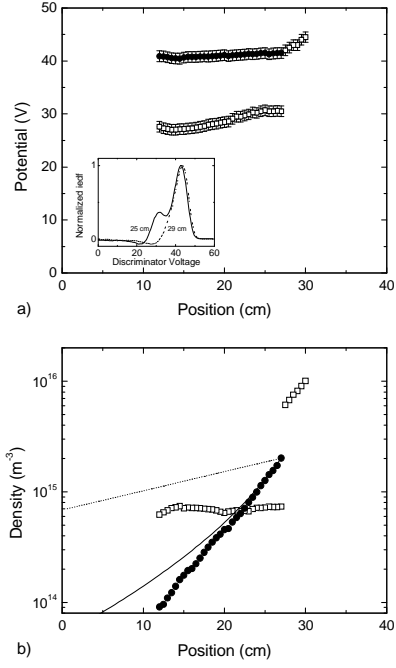}
\caption{\label{HDLTspatial}(a) Local plasma potential ($\square$) and beam potential ($\bullet$) as a function of position, normalized IEDF are plotted in the insert and (b) local ion density ($\square$) and beam ion density ($\bullet$) as a function of position, and theoretical spatial evolution assuming a flat profile (dashed line) and a hemispherical profile (solid line) for the electropositive DL. 0.17 mTorr, 90 G, 250 W.}
\end{figure}

Figure~\ref{HDLTspatial} shows results obtained at 0.17 mTorr, 90 G, 250 W as a function of axial
position. For positions below 28 cm, the IEDF clearly shows two peaks when the RFEA is facing the source while one
single peak is present above that axial position. When the RFEA is rotated by 90$^\circ$, only one peak is observed, centered at the same potential as the first peak when the RFEA is facing the source. This indicates
the presence of an abrupt increase of the plasma potential at z $\approx$ 28 cm, associated with a
dramatic increase in the plasma density, i.e. the double layer. This DL is associated with the presence of an ion beam in the low potential region, the speed of which is determined by the potential drop across the DL. The plasma
potential and plasma density are roughly constant below 28 cm, with the presence of a
beam at 42 V, the amplitude of which decreases exponentially when moving away from the DL; above 28
cm, one single peak at about 42 V is obtained, with a local ion density 10 times higher than
downstream (excluding the ion beam density), and increasing when moving up into the source tube.  The plasma density is determined using an electron temperature assumed to
be spatially homogeneous, which was measured to be 5$\pm$0.5 eV at 0.17 mTorr in the diffusion
chamber. With the voltage difference between the beam and local plasma component being 13
eV, the beam velocity is $
\sqrt{5}v_{\rm Bohm} \approx 2.2 v_{\rm Bohm}$ according to equation~\ref{eq.beamspeed}.
The beam measured downstream of the DL is therefore supersonic.

It should be noted that only indirect measurement of the DL was performed (via interpretation of the IEDFs and beam measurements). It was not possible to measure the plasma potential discontinuity by using the Langmuir probe. The probe disturbed the plasma such that the matching had to be dramatically changed when crossing the DL (the mismatch could reach 7 \% of the coupled power when moving the probe by only 1 cm a few centimeters when positioned downstream of the DL). 

\begin{figure}
\includegraphics[width = 0.8\linewidth]{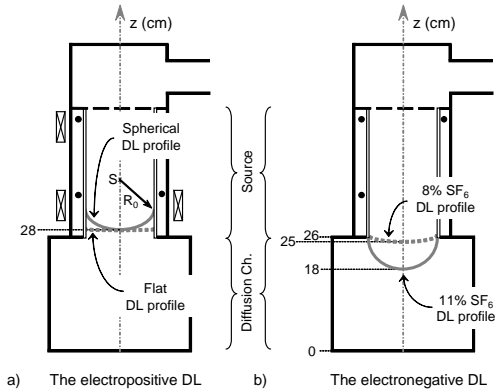}
\caption{\label{DLposition} (a) Schematic of the electropositive DL profiles chosen to fit the experimental data and (b) modification of the electronegative DL profile as a function of SF$_6$ concentration.}
\end{figure}

These features are very similar to the work presented in references~\cite{
Charl03a,Charl04c,Charl04b,Charl04,Charl05,Charl05a,Suthe05,Cohen03,Sun04,Sun05b,Biloi05, Keese05}. The most significant difference from the above references is the spatial variation of the ion beam flux in the downstream region. The main physical process involved in the damping of the ion beam is ion-neutral charge exchange collisions. The charge exchange mean free path for thermal ions in low pressure plasmas (the ion temperature being around 0.1 eV) is $\lambda_{i\ [\text{cm}]} = 3/p $ where $p$ is the pressure in mTorr~\cite{Liebe05}. The velocity dependence of the charge exchange cross section~\cite{Liebe05} leads to introducing a correction factor for the mean free path for a 15 eV beam and to estimating the effective mean free path as $\lambda  = \lambda_i/0.7$~\cite{Liebe06} (at 0.17 mTorr, $\lambda = 25 \text{cm}$). Although the actual geometry is a complex cylindrical geometry, we assume a one dimensional descripion of the ion beam density for the sake of simplicity. In order to compare the experimental beam decay to models,  assumptions on the actual DL 3-D profile ar required since direct measurement of the profile is not easily achieved. We therefore consider two cases, namely a flat profile (perpendicular to the $z$ axis) following the statements given in~\cite{Charl04b} and a hemi-spherical profile. These profiles are sketched in figure~\ref{DLposition}(a).  Assuming a flat profile, the one dimensional  ion beam flux ($\Gamma_{+b}$) conservation is 
$$\Gamma_{+b}' = -\Gamma_{+b}/\lambda $$ where the prime denotes the first derivative with respect to $z$. Integration of this equation with $Z_0 = 28\ \text{cm}$ leads to an ion beam density spatial evolution $n_{+b}(z) = n_{+b0}\exp(-(Z_0-z)/\lambda)$, plotted as a dashed line in figure~\ref{HDLTspatial}(b).   
The agreement with the experimental values is not satisfactory. The uncertainty over the exact pressure value, discussed in the following, could lead to a lower actual mean free path. Another reason may be due to the actual 3-D profile of the DL. Assuming the DL has a spherical shape, the ion beam flux conservation in spherical coordinates is
$$r^{-2}(r^2\Gamma_{+b})' = -\Gamma_{+b}/\lambda$$
where the prime denotes the first derivative with respect to $r$ (the origin of the spherical coordinates being at point S in figure~\ref{DLposition}(a)). Integration of this equation from the DL position assuming a $R_0$ radius  for the DL profile leads  to an ion beam density spatial evolution 
\begin{equation}
n_{+b}(r) = n_{+b0}\frac{R_0^2}{r^2}\exp{\left(-\frac{r-R_0}{\lambda}\right)}
\end{equation}
which can be expressed as a function of $z$
\begin{equation}
n_{+b}(z) = n_{+b0}\frac{R_0^2}{(R_0+Z_0-z)^2}\exp{\left(-\frac{Z_0-z}{\lambda}\right)}
\end{equation}
This theoretical evolution is plotted on figure~\ref{HDLTspatial}(b) as a solid line, and agrees reasonably well with the experimental values, especially above 20 cm.

 Understanding the spatial damping of the ion beam is important if this device is to be used as a space thruster as was proposed earlier~\cite{Charl03a,Gesto06}. Further investigations on the 3-D profile of the double layer is needed and optical diagnostics could provide useful informations if the experiment could be reproduced in a device with full optical access to the source tube. Another important issue is the actual value of the charge exchange mean free path.; precise knowledge of the operating pressure is thus needed. The operating pressure measured is the outer region of the diffusion chamber, using a 100 mTorr range Baratron pressure gauge. The relative error in the range 0.1-1 mTorr is the Baratron measurement could be as high as 50\%, even though absolute pressure values for DL existence are similar to the one obtained by Charles in~\cite{Charl05}.

\subsection{Magnetic field and pressure dependence}

Similarly to previous studies led at ANU~\cite{Charl03a,Suthe05,Charl05}, and  at WVU and Princeton University~\cite{Sun04,Sun05b,Biloi05}, two important experimental control parameters were identified in the DL formation process: the magnetic field magnitude and the operating pressure. This subsection presents a study of the DL sensitivity to these parameters.

\begin{figure}
\includegraphics[width = 0.7\linewidth]{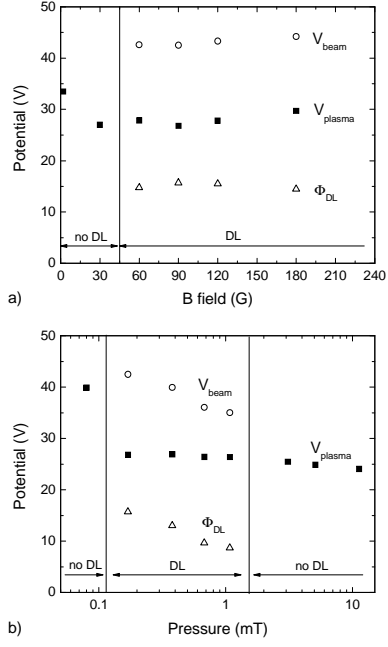}
\caption{\label{ParameterStudy} Local plasma potential, beam potential and  potential drop across the electropositive DL (a) as a function of the static magnetic field amplitude and (b) as a function of the pressure with the RFEA at z = 20 cm.}
\end{figure}

The first observation is that a sufficiently high and diverging static magnetic field is required to form a DL. Figure~\ref{ParameterStudy}(a) shows the behavior of the DL as a function of the magnetic field strength at a pressure of 0.17 mTorr, and a power of 250 W.
  For magnetic field magnitudes below
45 G, the IEDF shows only one peak, with a fairly high plasma density ($10^{16} \text{m}^{-3}$). Above
45 G, the double layer is formed. While the plasma density decreases slightly with increasing magnetic field (from $3\ 10^{15}\text{m}^{-3}$ at 45 G to $1\ 10^{15}\text{m}^{-3}$ at 90 G),
the plasma potential increases slightly from 28 V at 45 G to 30 V at 180 G.
The  potential drop across the DL (around 15 V for these conditions) varies weakly 
with the magnetic field magnitude; these variations remain however within the error bars of both the current measurement and the systematic IEDF processing. The beam density variations  appears to
follow the local plasma density behavior. 
As in  previous studies, the DL characteristics are fairly independent of the magnetic field above a critical value. However the actual shape and amplitude of the DL seems to be related to the actual geometry of the system and magnetic field topology.  Topological effects might explain the discrepancy between the high   potential drop across the DL observed by Charles~\cite{Charl05} and the lower potential drop obtained in this work and in Ref.~\cite{Sun04}. The DL position also depends on the geometry of the reactor, but remains close to the source - diffusion chamber junction (slightly downstream in WOMBAT~\cite{Suthe05}, and slightly upstream in Chi-Kung~\cite{Charl03a}, HELIX~\cite{Biloi05} and our source). These experiments show that above the critical magnetic field at which the DL forms, the DL characteristics do not depend on the magnetic field magnitude; thus the topology of the magnetic field might be the key parameter. The position of the DL in all these systems is close to the position of the maximum in the magnetic field gradient.

The second observation is that the double layer only appears at sufficiently low operating pressures. Figure~\ref{ParameterStudy}(b) shows the influence of the pressure on the DL characteristics for a 90 G magnetic field and
an operating power of 250 W with the RFEA positioned at z = 20 cm. 
No discharge could be sustained below 0.08 mTorr and between 0.08 and
0.1 mTorr, the measured IEDFs show only one peak, i.e. no DL is present.
Between 0.1 and 1 mTorr, two peaks are clearly observed, providing evidence of a DL.
When increasing the pressure, the plasma potential
downstream of the DL remains constant, while the upstream potential decreases. 
The
DL potential drop decreases dramatically with increasing pressure, as was determined in the model by Lieberman \textit{et al.}~\cite{Liebe06}. The beam flux  also decreases
with increasing pressure due to the decrease in acceleration from the
DL potential drop and the increase in collision frequency. The plasma density scales as expected at constant power:
increasing with increasing pressure, since the electron temperature required to sustain the
discharge decreases. 
At pressures above 1 mTorr the two peaks start to overlap.
 The resolution of the RFEA does not allow to distinguish between an upper pressure threshold  for DL formation or a gradual decay of the DL potential drop to zero. 
In any cases,  the DL disappeared at 3 mTorr.

The DL dependence on pressure and magnetic field strength are similar to previous experimental investigations and are correctly caught by the models developed in~\cite{Liebe06} and~\cite{Chen06}. Chen based his model on the hypothesis of a plasma frozen to the magnetic field lines~\cite{Chen06}. The magnetic confinement can be estimated from the value of the ratio $\omega_{ce}/\nu_m$~\cite{Liebe05}, where $\omega_{ce}$ is the electron cyclotron pulsation (proportional to the magnetic field magnitude) and $\nu_m$ is the total electron collision frequency (proportional to the pressure). Thus the assumption of a plasma frozen to the field lines is satisfied at high magnetic field and/or at low pressure, which are the conditions for DL occurrence. In this model, the  potential drop across the DL  is dependent on the electron temperature and the condition for development of a non-neutral  region is {for} the plasma column radius to  increase by 28\%. An analysis of the calculated magnetic field lines (assuming a plasma frozen to the field lines) leads to a DL position at z $\approx$ 30 cm according to this condition, which is close to the observed value (28 cm) and to the maximum of the magnetic field gradient.
The model by Lieberman \textit{et al.}~\cite{Liebe06} explicitly accounts for the pressure dependence of the DL. All trends emphasized in this study are caught by the model, namely the pressure dependence of the  potential drop across the DL , a low pressure limit for DL existence due to a  low ionization term from accelerated electrons  upstream, and a high pressure limit when no ionization is needed from  accelerated electrons  upstream. However, the model in~\cite{Liebe06} does not include a magnetic field, which is a key parameter of this experiment.

The experimental investigations also show that the  potential drop across the DL  was fairly insensitive to the coupled power. However the ion plasma density to ion beam density ratio in the diffusion chamber is highly dependent on the power, increasing with increasing power~\cite{Pliho05b}.

\subsection{Ignition of the electropositive double layer}
\begin{figure}
\includegraphics[width = 0.65\linewidth]{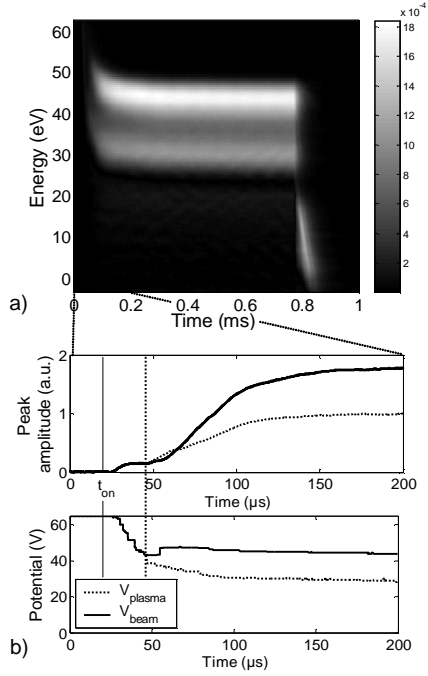}
\caption{\label{IEDFTimeresolved}(a) Time resolved IEDF when the plasma is pulsed and (b) the associated beam and plasma potentials and amplitudes as a function of time (zoom of the ignition phase) showing ignition of the electropositive DL.}
\end{figure}

Using the time resolved data acquisition system, the ignition and the time development of the DL was measured. The discharge was operated in a pulsed mode, with a 750 $\mu$s 'on' period and 500 $\mu$s 'off' period. Although the lifetime of some Ar metastables states is of the order of hundreds of microseconds at the current operating pressures, the 'off' period allows sufficient time to destroy all charged species. Time-resolved IEDFs are
shown in figure~\ref{IEDFTimeresolved}(a) but need to be analyzed bearing in mind the generator time scale.
Analysis of the pulsed mode operation of the radiofrequency power generator on a 50 $\Omega$ load showed that the typical generator rise-time is 150 $\mu$s. The transient plasma operation time will thus last at least 150 $\mu$s.

The IEDFs are processed to follow
the amplitude and potential of the two peaks, which are given in
figure~\ref{IEDFTimeresolved}(b). The beginning of the 'on' time is located at
t = 20 $\mu$s, and only one peak is present before t = 50$\mu$s.  The beam potential reaches its steady-state value over 10-20 $\mu$s, while the
downstream plasma potential keeps on decreasing (from 38 to 30 V)
over the next 50 $\mu$s. Formation of the double layer occurs less than 30 $\mu$s after plasma break down and the steady state potential values are reached well before steady state density is reached. The amplitude of both the beam component
and local plasma component increase during the first 150
$\mu$s of operation, and increasing at the same rate.

These observations confirm and complement Charles's investigations~\cite{Charl04a}, by providing a detailed analysis of the IEDF temporal evolution. The time scale for the DL to develop is found in these experiments to be around 100 $\mu$s. However,in these experiments, an exact evaluation of the DL time development was not possible since the ignition time is less than the generator risetime (the overshoot in figure 4 of reference~\cite{Charl04a} might be due to the generator temporal dynamics). Temporal evolution of the downstream beam velocity from LIF in the HELIX system~\cite{Biloi05} over long pulses showed that a steady state was not reached after 100 ms of operation.  In our system, it may be that the DL forms and develops rapidly over the first 200 $\mu$s, and then experiences a slower rate of change as was observed in the HELIX reactor. In our system, it was not possible to operate with longer pulses so as to observe a change in the temporal dynamics of the DL on large time scales (hundreds of ms to seconds).

\subsection{Influence of the boundary
conditions}\label{sec.boundary}

\begin{figure}
\includegraphics[width = 0.7\linewidth]{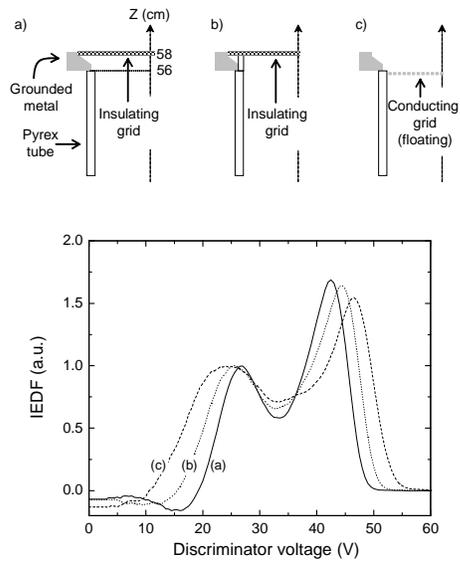}
\caption{\label{SchemaCLESA} Boundary conditions terminating the plasma studied and associated IEDF measured in the diffusion chamber, downstream of the electropositive DL.}
\end{figure}

We have investigated the influence of changing  the boundary
conditions at the end of the source tube. 
Three different configurations were
experimentally studied: (a) an insulating grid positioned 2 cm above the top of the pyrex source tube with a dc grounded metal cylinder in-between, (b) an insulating grid 2 cm above the top of the pyrex source tube with an insulating pyrex cylinder in-between and (c) a non-dc grounded conducting grid at the top of the pyrex source tube. These configurations are sketched in figure~\ref{SchemaCLESA}. All results presented in the preceding subsections and in the following section were acquired with  condition (a), namely an insulating grid at the end of the pyrex tube
with a piece of grounded metal.

The bottom part of figure~\ref{SchemaCLESA} shows IEDFs normalized to one at the plasma potential, when the RFEA is
positioned at z = 20cm, for a 0.17 mTorr, 90 G, 250 W plasma. The variation of the ion
saturation flux is  within  3\% for the three configurations presented. The beam amplitudes decrease by a few percent (8\% from geometry (a)
to geometry (c)). A noticeable variation of the potentials is
observed: the local plasma potential decreases: 26.5 V, 25.5 V
and 24.5 V, while the beam potential increases: 42.5 V, 44.2
V, 46.5 V, for geometry (a), (b) and (c) respectively. The
broadening of the low potential peaks increases slightly when scanning from
geometry (a) to geometry (c). Due to the low operating pressure and large mean free path, the RFEA collecting length is large (approximately one mean free path). The IEDF's first peak broadening could originate as a result of smaller plasma potential gradient lengths in the downstream region from geometry (a) to (c). A gradient length less than the mean free path leads to a non-mono-energetic beam and peak broadening. The potential drop across the DL changed noticeably with the boundary conditions, the
higher  potential drop across the DL being obtained for the floating conducting
grid. This investigation confirms the influence of the geometry, but these results are significantly different from those published in Ref.~\cite{Charl05} which showed that the beam component relative to the
plasma component could triple when changing the position of the
end plate by only 2 cm. 

Interestingly, the double layer was formed for all boundary
conditions under the same experimental conditions, including the partially dc grounded case (condition a). DLs with conditions (b) and (c) have to be current free, while DL with condition (a) could be current carrying. Whatever its nature (current-free or possibly current-carrying), the DL's characteristics are weakly modified as emphasized in Ref.~\cite{Liebe06}.
Experiments show that the same trends are maintained as a function of
pressure, magnetic field magnitude and position whatever the source boundary
conditions.

\section{Double layers in electronegative mixtures\label{sec.ENDL}}

Double layers have been observed in the same experimental device operating without a magnetic field  in electronegative gas mixtures~\cite{Pliho05,Pliho05a}. In this section, we present a detailed experimental investigation of the static electronegative DL introduced in~\cite{Pliho05}. Results were obtained in a stable inductive mode, i.e. far from the relaxation oscillations between the capacitive and inductive modes. The section is organized as follows: we first develop results obtained in Ar/SF$_6$ gas mixtures and then present results obtained in Ar/O$_2$ gas mixtures.

\subsection{Conditions for DL formation}

We emphasized in the introduction that DL formation was highly dependent on the SF$_6$ concentration: a DL forms at the interface between the source tube and the diffusion chamber when the SF$_6$ concentration is above a critical concentration. This DL remains static for a narrow range of SF$_6$ concentration (namely 8 to 13\% at 1.5 mTorr), before a propagating mode is established. 
Variation of the SF$_6$ concentration within the [8-13\%] window has a dramatic impact on the DL characteristics and overall plasma equilibrium. At a typical 1.5 mTorr plasma pressure, the position of the DL on axis is measured at z = 25 cm for a 8\% SF$_6$ concentration and as low as 18 cm for a 11\% SF$_6$ concentration. Direct measurement of the 3-D profile is not easily achieved. However from visual observation through an optical window in the diffusion chamber, the DL profile can be seen as an interface between a bright plasma in the source and a more diffuse plasma in the diffusion chamber. The actual 3-D profile has a hemi-spherical shape entering the diffusion chamber and is attached to the bottom of the source tube. This is similar to the plasma-sac previously described in Ref.~\cite{Andre71}. A change of the profile observed when increasing the SF$_6$ concentration is given schematically in figure~\ref{DLposition}(b). 
The plasma equilibrium is also strongly modified with a change in SF$_6$ concentration: the downstream plasma potential is decreasing from 18 to 15 V for SF$_6$ concentrations increasing from 8 to 11\%. The DL's characteristics are sensitive to any parameter change; slight changes in pressure or power  lead to a modification of the DL profile at a given SF$_6$ concentration. Over time (month to month) similar slight changes were observed  due to wall contamination or thermal drift for the same SF$_6$ concentration. However the same trends and values were reproduced over a time period of two years. All SF$_6$ concentrations specified in this paper are calibrated according to a reference average concentration.
Furthermore, intrusive probing of the plasma parameters can modify the DL position. At low plasma density (coupled power around 100 W), the DL formation was observed, but its shape could be modified by moving the Langmuir probe in its vicinity. This behavior was less extreme at higher plasma densities, since the probe area is small compared to the plasma section. 
Inserting the RFEA in the diffusion chamber changes the 3-D profile of the DL. When the RFEA is moved upward in the chamber, the DL is "attracted" towards the RFEA. It was not possible to measure the IEDF close to the DL without disturbing it: for a 8\% SF$_6$ concentration, the DL profile is not modified (crossing the $z$ axis at 25 cm) when the RFEA is at z = 8-9 cm, , while the RFEA destroys the DL when positioned at z = 15 cm.

All results presented in the following subsections were obtained slightly above the critical SF$_6$ concentration for DL formation, i.e. the DL profile is close to the 8\% SF$_6$ case plotted in figure~\ref{DLposition}. The reader should keep in mind the high sensitivity of the DL to intrusive probes, especially when analyzing RFEA results.

Electronegative DL formation was observed to be easier than electropositive DL formation. The electronegative static DL was created for all pressures in the range 0.5-6 mTorr, for all power from 50 to 1500 W (i.e. both in the capacitive and inductive power coupling modes). For any power-pressure range, a static electronegative DL forms within a narrow window of low SF$_6$ concentration.
At higher pressures (above 6 mTorr), the electronegative DL was always observed propagating.

\subsection{Spatial evolution of electron parameters\label{sec.ENDLspatial}}

\begin{figure}
\includegraphics[width = 0.5\linewidth]{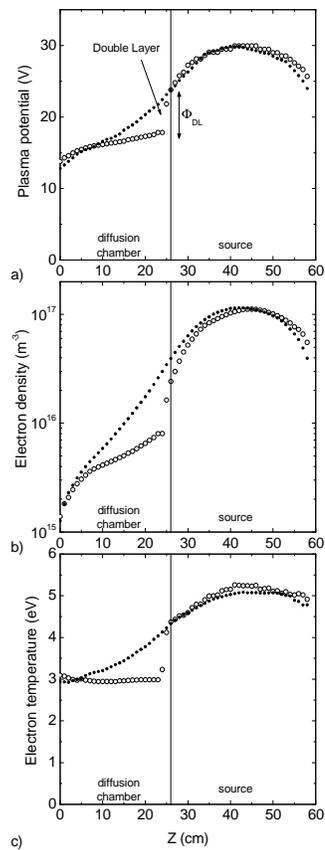}
\caption{\label{ENDLSpatial} Spatial variation of (a) the plasma potential, (b) the electron density and (c)  the electron temperature in Ar/SF$_6$ mixtures: 4\% SF$_6$ concentration (no DL - dots) and 8\% concentration (showing electronegative DL formation - open circles). 600W, 1.5 mTorr.}
\end{figure}

Figure~\ref{ENDLSpatial} shows the spatial evolution of the plasma parameters on axis for two SF$_6$ concentrations: 4\% (dots) and 8\% (open circles). The lower SF$_6$ concentration profiles correspond to the no DL case and are similar to pure argon profiles (when operated without a d.c. magnetic field); the plasma is created in the source and diffuses continuously into the diffusion chamber. The profiles for a 8\% concentration clearly show a discontinuity at the interface between the source and the diffusion chamber; the DL is created at this interface (seen at z $\approx$ 25cm for all plasma parameters). 
In the source, the plasma parameter profiles are similar in both the non-DL and DL cases. 
All results presented in this subsection and in subsections~\ref{sec.RFEAENDL}-~\ref{sec.ENDLphoto} were obtained for a 1.1 mTorr pressure measured when the discharge is "off". Due to dissociation of the molecular gas, the pressure lies in the range 1.4-1.5 mTorr when the plasma is ignited.

The  potential drop across the DL  is difficult to measure from the Langmuir probe axial profiles, since upstream of the DL a continuous decrease in potential, similar to a pre-sheath profile, is observed, leading to a somewhat indeterminate beginning of the non-neutral region. However this potential drop can be estimated to be around 7 V. The potential drop can also be estimated from equation 6 in Ref.~\cite{Biloi05} which estimates the DL amplitude from an electron flux balance between the upstream and downstream plasmas, and gives a value of 6.7 V (when assuming the upstream plasma edge at z = 25 cm).

\begin{figure}
\includegraphics[width = 0.65\linewidth]{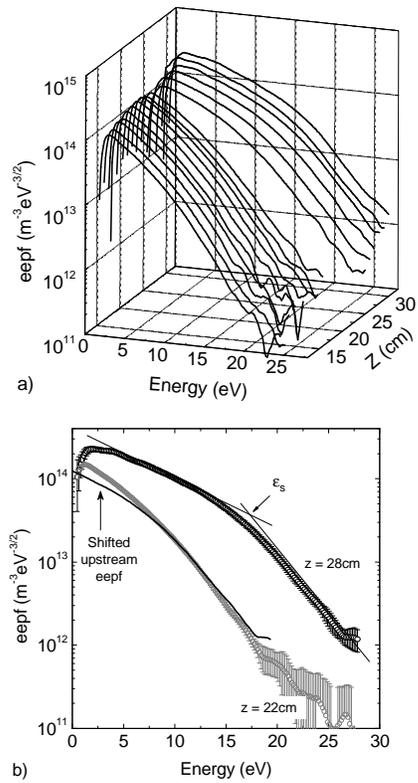}
\caption{\label{ENDLeepf} (a) The spatial variation of the eepf in the presence of the electronegative DL, and (b) the eepf upstream of the DL (z = 28 cm, black circles), downstream of the DL (z = 22 cm, gray circles) and shifted upstream eepf (solid black line).}
\end{figure}

Upstream and downstream of the DL, the electron dynamics is clearly different. Figure~\ref{ENDLeepf}(a) shows the eepf as a function of energy and axial position. The DL separates maxwellian-like distributions downstream and truncated-maxwellian like distributions upstream. Figure~\ref{ENDLeepf}(b) gives a view of eepfs at axial position 22 cm and 28 cm (namely 3 cm downstream and 3 cm upstream of the DL), plotted as symbols. The upstream distribution is similar to those obtained in low pressure inductively coupled plasmas~\cite{Godya02}. The truncature energy $\epsilon_s$ tends toward the sheath voltage (difference between the plasma potential and floating potential in our situation) at very low pressure (below 1 mTorr) and toward the excitation collision energy threshold (above 10 mTorr). In figure~\ref{ENDLeepf}(b), $\epsilon_s$ is estimated to be 17 V (slightly higher than the excitation collision energy threshold of 11.5 eV in argon).

The electron density and effective electron temperature determined from the eepf are consistent with the results presented in figure~\ref{ENDLSpatial}. The electron density obtained by integration of the eepf follows the same trend as the electron density computed from the I-V curve analysis with absolute value discrepancies less than 10\%. The spatial evolution of the effective electron temperature exhibits an abrupt drop at the DL position. However, the effective electron temperature values differ from those derived from direct I-V curves analysis: downstream of the DL the effective electron temperature is measured to be 3.8 - 4 eV, and 5.5 - 6 eV upstream. The abrupt difference in effective temperature is in the range 1.5 - 2 eV.

The change in the electron dynamics across the DL could originate in the shift of the upstream eepf through the retarding potential of the DL. A 8.5 V Boltzmann shift of the upstream eepf of figure~\ref{ENDLeepf}(b) is  represented by the black solid line. This shift cuts off the lower energy and brings the distribution into agreement with the downstream distribution.
The 8.5 V value is obtained from the difference in potential at z = 22 and z = 28 cm taken from figure~\ref{ENDLSpatial}(a). 
 However the shifted distribution is somewhat different from the downstream distribution at low energies. This could be due to distinct physical processes upstream and downstream (ionization, excitation and attachment processes may vary due to a difference in the plasma equilibrium and species). The electron density and effective electron temperature calculated from the shifted distribution are $8\times 10^{15}\text{cm}^{-3}$ and 3.9 eV respectively (compared to $9\times 10^{15}\text{cm}^{-3}$ and 3.85 eV from integration of the energy distribution at 22 cm).

\subsection{Evidence of a positive ion beam in the downstream region\label{sec.RFEAENDL}}

\begin{figure}	
\includegraphics[width = 0.7\linewidth]{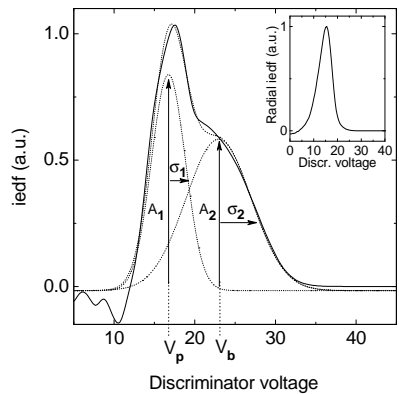}
\caption{\label{RFEAENDL} An experimentally measured IEDF showing two peaks and the associated fit used for systematic analysis of the electronegative DL. Insert: IEDF measured when the RFEA is in the radial position.}
\end{figure}

\begin{figure}
\includegraphics[width = 0.7\linewidth]{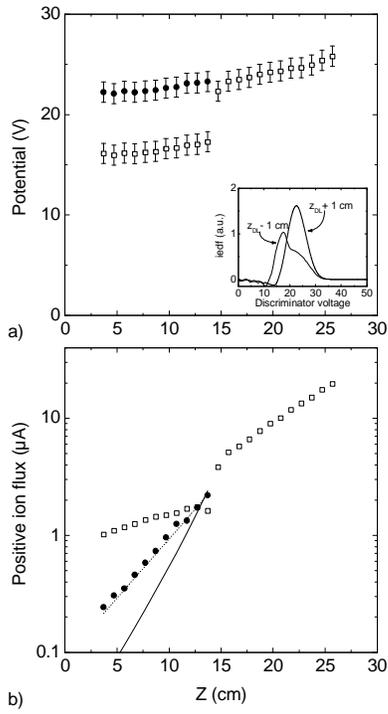}
\caption{\label{ENDLBeamSpatial}a) The local plasma potential ($\square$) and the beam potential ($\bullet$) as a function of position (IEDF are plotted in the insert) and (b) the local ion density ($\square$) and the beam ion density ($\bullet$) as a function of position, and the theoretical spatial evolution assuming a flat profile (dashed line) and a hemispherical profile (solid line) for the electronegative DL. 600 W, 1.5 mTorr.}
\end{figure}

Similarly to the electropositive case, IEDFs were measured in the low potential region with the RFEA facing the source. Experimental evidence of an ion beam in the diffusion chamber, accelerated through the DL, is drawn from the IEDFs. Figure~\ref{RFEAENDL} shows an experimental IEDF downstream of the DL exhibiting two peaks. Due to the RFEA's resolution, these two peaks are not clearly separated. 
However, as shown in figure~\ref{RFEAENDL}, the IEDF can be fitted as the sum of two gaussian functions (amplitude $A_i$ and deviation $\sigma_i$). Following the analysis presented previously for the electropositive DL (section III), the IEDF provides a measure of the plasma potential ($V_p$), the beam speed ($V_b-V_p$) and the positive ion flux from the local plasma ($\Gamma_+ A_1\sigma_1/(A_1\sigma_1+A_2\sigma_2)$) and the beam flux ($\Gamma_+ A_2\sigma_2/(A_1\sigma_1+A_2\sigma_2)$). The beam speed is an indirect measurement of the  potential drop within the DL. The dispersion $\sigma_1$ of the gaussian centered at the plasma potential is smaller than the dispersion  $\sigma_2$ of the gaussian centered at the beam potential. The dispersion of the peak centered at the plasma potential is due to the resolution of the RFEA and is not a measure of the ion temperature since $\sigma_1$ is 3 V, which is far higher than the typical positive ion temperature in such a plasma (which is less than 0.5 eV). Concerning the higher value of $\sigma_2$, it is possibme that the ion beam is not mono-energetic and spreads out, due to the gradient in the plasma potential upstream of the DL (as given in figure~\ref{ENDLSpatial}(a)), whose gradient length is of the same order as the ion mean free path. 
The insert of figure~\ref{ENDLBeamSpatial} shows the IEDF acquired with the RFEA inserted radially, into the mid-plane of the diffusion chamber at a 6 cm radial extension from the axis. This IEDF has only one peak, centered at the plasma potential that was determined from the axial IEDF. The second peak observed on the axial IEDF is therefore due to a population created upstream and accelerated within the DL.

Measuring the IEDF downstream is a non-invasive probing method of the DL existence. However, we previously emphasized that when the RFEA is positioned in the upper part of the diffusion chamber, the DL profile was profoundly affected.
 Keeping this limitation in mind,
 figure~\ref{ENDLBeamSpatial} shows the spatial evolution of the plasma potential and the beam potential, as well as the local and beam fluxes from IEDF measurements. The DL position is lowered down to z = 15 cm for the higher probing positions.  
 Assuming the damping of the ion beam flux to originate in ion-neutral charge exchange, the mean free path for this process is taken to be, in a similar fashion to the electropositive case, $\lambda_{[cm]} = 3/(0.7p_{[mTorr]})$ where $p$ is the argon partial pressure. The argon pressure was set to be 1 mTorr when the plasma was off. Even though the pressure is higher when the plasma is struck, we state that this increase in pressure is due to SF$_6$ molecular dissociation, and that the argon partial pressure remains at 1 mTorr, leading to $\lambda =  4.3\ \text{cm}$. As is similar for the electropositive case, theoretical evolutions of the beam damping are compared to the measured beam amplitudes in figure~\ref{ENDLBeamSpatial}(b) for two assumed DL profiles. The flat profile case is plotted as a dashed line, while the hemi-spherical profile case (with $R_0$ = 10 cm) is plotted as a solid line.

Interestingly, the best fit is obtained for an assumed flat profile of the DL, although observations clearly indicate an experimental hemi-spherical profile (with a curvature radius in the range 12-15 cm). The effect of the RFEA disturbances were not taken into account in these theoretical plots. A decrease of the DL position by the RFEA attraction leads to a reduced effective mean free path. On the other hand, deeper penetration of the DL into the diffusion chamber leads to a reduction of the ion flux crossing the DL (since the total flux remains constant) and this effect could compensate for the previous one. Damping of the beam is thus weakly understood and is theoretically underestimated for the electropositive DL and overestimated for the electronegative DL. Precise measurement of the ion beam damping length can only be obtained by the use of non-intrusive Doppler-Fizeau shifted laser diagnostics.

\subsection{Negative ion fraction measurements\label{sec.ENDLphoto}}

\begin{figure}
\includegraphics[width = 0.7\linewidth]{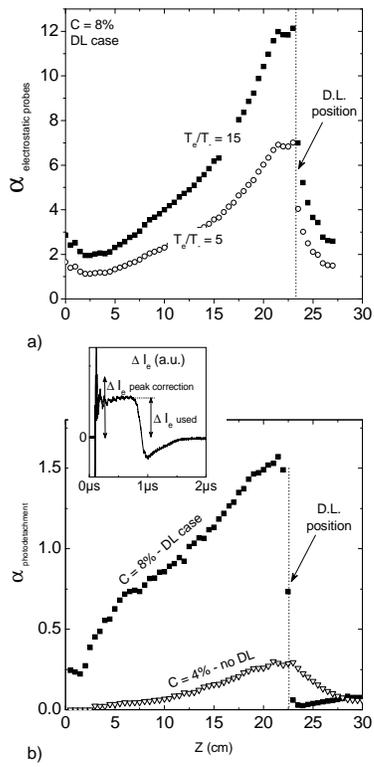}
\caption{\label{PhotodetSF6}  The spatial evolution of the negative ion fraction measured (a) by the two probes technique for two negative ion temperature estimates and (b) by aser induced photodetachment (LIP) for two SF$_6$ concentrations. The insert in figure (b) shows the increase of the electron current due to LIP.}
\end{figure}

Spatial evolution of the negative ion fraction in the diffusion chamber have been presented in reference~\cite{Pliho05}. The evolution was processed from electrostatic probe measurements according to the two probe technique described in~\cite{Chabe99}. This technique relies on a sheath theory~\cite{Sheri99a} where negative ions are assumed to be in Boltzmann equilibrium (the electric force is balanced by the thermodynamic pressure which is proportional to  the actual negative ion temperature) as is usually the case in low pressure plasmas~\cite{Frank00b}. It requires the knowledge of (i) the dominant positive ion mass, and (ii) the negative ion temperature. The negative ion temperature is assumed to be proportional to the electron temperature, with a coefficient $\gamma=T_e/T_-$ usually estimated in the range 10-20 in similar discharges~\cite{Chabe99}. Moreover, the resolution of the technique is $\gamma$-dependent~\cite{Chabe99,Sheri99a}, but is qualitatively limited to $\alpha>2.5-3$ in similar plasmas. The results presented in~\cite{Pliho05} were processed assuming $\gamma = 15$ and an effective positive ion mass of 40 a.m.u. These results are reproduced in figure~\ref{PhotodetSF6}(a). According to this technique, the  negative ion fraction scales as the square root of $\gamma$ and the effect of using $\gamma = 5$ is plotted in figure~\ref{PhotodetSF6}(a)(open circles). The limitations of these results should be highlighted  before comparison with LIP measurements. 
The first limitation lies in the negative ion temperature estimate, and the validity of a Boltzmann relation for negative ions. The second limitation is the resolution of the technique: we are not able to measure a negative ion fraction lower than 3 for $\gamma = 15$. Hence, the actual negative ion fraction value might be in the range [0;3] when a value below 3 is measured. Negative ion fractions upstream of the DL and in the lower part of the diffusion chamber are therefore upper bounds. The last limitation is due to the ion beam existence in the downstream plasma. The planar probe used is facing the source and the positive ion saturation flux has two components: a local flux and a beam flux. The sheath theory does not apply in the presence of a beam. Since electroneutrality reads $n_{+p}+n_{+b} = (1+\alpha)n_e$ where $n_{+p}$ is the local ion density and $n_{+b}$ the beam ion density, one should use $\Gamma_{+p}+\Gamma_{+b}\frac{u_B}{v_b}$ as the positive ion flux in the negative ion calculation, where $u_B$ is the modified Bohm velocity and $v_b$ the beam velocity. Since $v_b\gg u_B$ if the negative ion fraction is above 1, a correction can be estimated by subtracting the beam flux from the total flux. Within the first 4 cm downstream of the DL, this leads to a correction  factor 0.5 according to the RFEA's flux measurements shown in figure~\ref{ENDLBeamSpatial}(b). We thus believe that a few centimeters downstream of the DL the actual value of the negative ion fraction from electrostatic probes could be half the values plotted in figure~\ref{PhotodetSF6}(a).

Probe based laser induced photodetachment (LIP) allows the direct measurement of the negative ion fraction as was outlined in section~\ref{sec.exp}.
Figure~\ref{PhotodetSF6}(b) presents the spatial evolution of the negative ion fraction in the diffusion chamber and in the bottom of the source for the two SF$_6$ concentrations (4 and 8\%) previously detailed. Negative ion fractions measured by the two probe technique and displayed in figure~\ref{PhotodetSF6}(a) are for a 8\% SF$_6$ concentration  (as was emphasized at the beginning of the section, the exact position of the DL is highly sensitive to any parameter change, and is observed to have varied by a few millimeters between electrostatic probes and LIP measurements). Although LIP is efficient for negative ion fraction measurements in presence of an ion beam, this technique also has limitations. All necessary experimental conditions were verified for all probe positions in the diffusion chamber (saturation of the collected current, colinearity of the probe tip with beam axis); however one should estimate the relative error to be around 15\%.
Another concern is the measurement of high negative ion fractions (typically $n_-/n_e>1$). A typical temporal profile for the increase of the probe current ($\Delta I_e$) during photodetachment  is shown in the insert of figure~\ref{PhotodetSF6}(b). The saturation value of the probe current ($\Delta I_e\ \text{used}$) was used for the determination of $\alpha$ according to equation~\ref{eq.alpha}. However, according to Ref.~\cite{El96}, a sharp peak may be observed at the beginning of the probe current pulse for large negative ion fraction. In this case the authors in Ref.~\cite{El96} recommend to use the peak value ($\Delta I_e\ \text{peak correction}$) in equation~\ref{eq.alpha}. Our signals show an overshoot at the beginning of the pulse and large high frequency oscillations are also observed. Since we are uncertain as to the origin of the high frequency oscillations, we ignored the peak correction. Taking this peak correction into account would lead to a 50\% increase in the negative ion fraction plotted in figure~\ref{PhotodetSF6}(b). 
Finally, one further limitation of the measurement is the uncertainty with the SF$_5$ electron affinity, and the fact that the laser's photons at 355 nm  may not be able to detach SF$_5^-$. An upper estimate of the SF$_5^-$ fraction is 0.5 (Kono estimates this fraction to be a few hundredths  in an ICP~\cite{Kono02} and was measured in another device  to be 0.15~\cite{Sasak97}). Thus, it appears that the actual negative ion fraction from the LIP technique lies between the measured value presented in figure~\ref{PhotodetSF6}(b) and twice this value.  

Although no definitive conclusions on the absolute values of the negative ion fraction could be drawn, the spatial evolution is clearly identified. The negative ion fraction experiences an abrupt increase at the DL position when entering the low plasma potential region. The DL separates a highly electronegative plasma in the diffusion chamber from an electropositive plasma in the source. Downstream of the DL, the negative ion fraction gradually decreases  from its highest value, at the DL position, to a very small value at the bottom of the chamber. This spatial evolution was captured in a recent model~\cite{Chabe06} and seems to be a result of the plasma expansion. In the no-DL case, the negative ion fraction profile displays a shallow maximum at the two chambers interface. However, either side of this maximum, the negative ion fraction decreases gradually.

We believe that the value of the negative ion fraction measured slightly above the DL is actually zero. This is confirmed by the negative ion fraction measured by LIP in O$_2$ presented in the following section. 

Though absolute values of the negative ion fraction published in Ref.~\cite{Pliho05} might be overestimated, LIP confirms that the DL is an internal sheath between a highly electronegative plasma ($\alpha > 1$) in the low potential region and a weakly electronegative  plasma ($\alpha < 1$) in the high potential region. Note that a weakly electronegative plasma has an electropositive pre-sheath and the negative ion fraction could be as high as 1 in the center of the source. Unfortunately our experimental arrangement did not allow us to measure the negative ion fraction in the source.

\subsection{Parametric dependences\label{sec.ENDLparam}}

\begin{figure}\includegraphics[width = 0.65\linewidth]{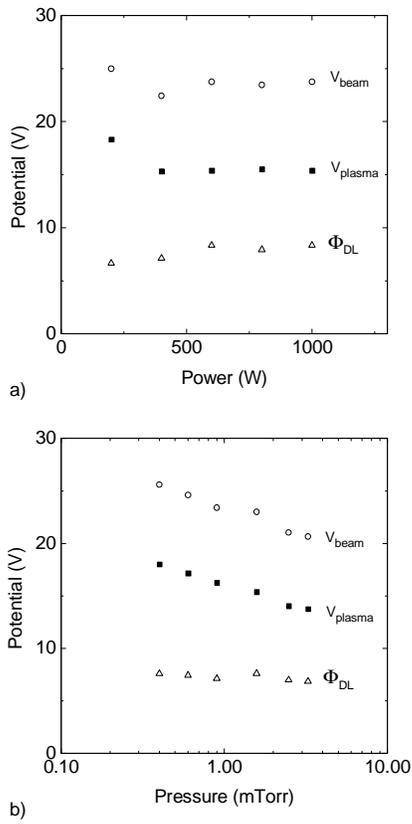}
\caption{\label{ENDLparam}  The evolution of the plasma and beam potentials as a function of (a) power at 1.5 mTorr, and (b) pressure at 600 W, with the RFEA at z = 10 cm.}
\end{figure}

The DL characteristics were investigated as a function of the external parameters. Figure~\ref{ENDLparam}(a) shows the power dependence of the potential obtained from the IEDF when operating at 1 mTorr, while maintaining the SF$_6$ concentration at 8$\pm$0.5\% such that the DL position is at 25$\pm$1 cm. Figure~\ref{ENDLparam}(b) shows the same analysis as a function of pressure in the range 0.5-6 mTorr while the power coupled to the plasma is kept constant at 600 W.
Variation of the coupled power does not influence the  potential drop across the DL, which is consistently measured to be 8$\pm$1 V from the IEDF analysis. 
The influence of pressure is somewhat surprising. DLs were observed for a broad range of pressure, from 0.5 mTorr to 10 mTorr - however above 6 mTorr, static DLs were only observed at very high power (above 2000 W) and propagating DLs observed at low power. As the pressure is increased, the particle balance imposes that the electron temperature decreases. IEDFs analysis shows a decrease of both the plasma potential and the beam potential while the  potential drop across the DL remains at 8$\pm$1 V when varying the pressure from 0.5 to 6 mTorr. An important observation is the pressure dependence of the SF$_6$ concentration for DL formation: decreasing from 15\% at 0.5 mTorr to 5.5\% at 6 mTorr. The independence of the electronegative DL potential drop with pressure is distinct from the electropositive case, whose potential drop has been shown to be proportional to the electron temperature.

As in the electropositive magnetized case, the influence of changing the boundary conditions has been investigated. We observed that for the same gas mixture, pressure and power, a change in boundary conditions does not produce a significant change in the plasma potential profile. From geometry (a) to geometry (c) (presented on figure~\ref{SchemaCLESA}), the plasma potential has the same spatial evolution, with a 2 V offset for geometry (c). The  potential drop across the DL  and the DL position was not modified by varying the boundary conditions. The formation of a DL in geometry (c) attached to the source tube forces the DL to be current-free, and we believe that any of the three geometries will lead to a current-free DL.

\subsection{The Ar/O$_2$ mixture}

\begin{figure}
\includegraphics[width = 0.7\linewidth]{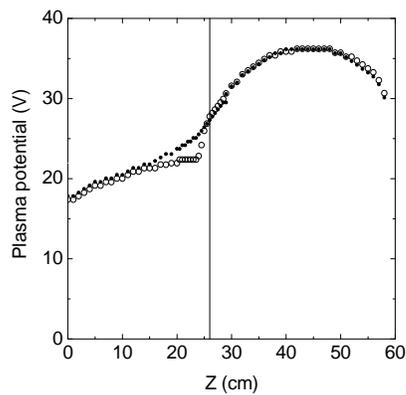}
\caption{\label{ENDLSpatialO2} The spatial variation of the plasma potential in Ar/O$_2$ mixtures: 68\% O$_2$ (no DL - dots), 72\% O$_2$ (electronegative DL case - open circles).}
\end{figure}

\begin{figure}
\includegraphics[width = 0.7\linewidth]{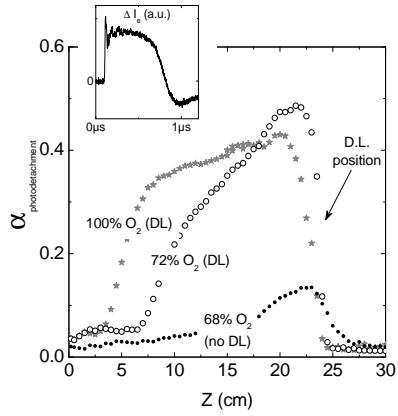}
\caption{\label{PhotodetO2}  The spatial evolution of the negative ion fraction from LIP for  a 68\% O$_2$ mixture (no-DL) and 72\% O$_2$ and pure O$_2$ mixtures (electronegative DL). The insert shows the increase of the electron current due to laser induced photodetachment.}
\end{figure}

DLs were also observed when operating in Ar/O$_2$ gas mixture. As for the Ar/SF$_6$ gas mixture, the DL forms when the O$_2$ concentration is above a critical threshold. Figure~\ref{ENDLSpatialO2} shows the spatial evolution of the plasma potential  for 68\% (dots) and 72\% (open circles) O$_2$ concentrations showing the DL formation at an operating pressure of  1 mTorr. For any O$_2$ concentration when diluted in argon, the plasma remains stationary, and the discharge does not enter the propagating DL regime, which is the main difference from the Ar/SF$_6$ mixture. The characteristics of the DL are similar to the Ar/SF$_6$ case: the DL separates (i) a high plasma potential region in the source and a low plasma potential region in the diffusion chamber, (ii) a high electron density plasma upstream and a low electron density plasma downstream, and (iii) a high electron temperature plasma upstream and a low electron  temperature plasma downstream. LIP measurements of the negative ion fraction in Ar/O$_2$ gas mixtures is simplified în comparison to Ar/SF$_6$ gas mixtures: all negative ions created (O$^-$, O$_2^-$ and O$_3^-$ ) can be photodetached by 532 nm photons. Thus, the measurements do not present a large uncertainty in the absolute values. The laser beam energy dependence of the LIP signals provide a measurement of the photodetachment cross section and therefore of the dominant negative ion. For the present operating conditions (1 mTorr - 750 W), the dominant negative ion is O$^-$ as is commonly observed in similar discharges~\cite{Kimur01,Corr03}.  Figure~\ref{PhotodetO2} shows the axial profile of the negative ion fraction in the diffusion chamber for three O$_2$ concentrations (68, 72 and 100\%).
Strong similarities with Ar/SF$_6$ gas mixtures are observed; the negative ion fraction profile is continuous with low absolute values for the non DL case, and for the DL case an abrupt drop is observed at the DL position. Absolute values of the negative ion fraction downstream of the DL are lower than in Ar/SF$_6$ gas mixtures due to lower O$_2$ attachment coefficients.
These profiles show that the negative ion fraction is close to zero in the source region, which is consistent with the results in Ar/SF$_6$ mixtures. The negative ion fraction also experiences a decrease at the bottom of the diffusion chamber, which is typical of stratification in electronegative plasmas~\cite{Frank92,Licht94,Sheri99a}. If we were to take into account the peak at the beginning of the pulse, the negative ion fraction would be increased by 20\% (see insert in figure~\ref{PhotodetO2}).

\section{Conclusion}

An extended experimental study of  static DLs in a device composed of a source chamber attached to a larger expanding chamber has been presented. 
Positive ions created upstream are accelerated across the DL, thus leading to a positive ion beam in the downstream region. The positive ion beam could be utilized as a long term high power space electric propulsion system as was proposed by Charles~\cite{Charl03a}.

With an electropositive gas, the DL forms at a very low pressure if a highly diverging magnetic field forces a strong plasma expansion. The typical voltage drop across this electropositive DL is 20 V at 0.17 mTorr. Several boundary conditions were investigated at the end of the source tube and showed that the DLs are not necessarily current-free. They are strongly attached to the magnetic field lines and appeared to be very stable. In particular their position was not modified by the intrusive electrical diagnostics approaching from downstream, but crossing the DL with Langmuir probes would destroy it.

DLs were also formed when using electronegative gas mixtures (Ar/SF$_6$ and Ar/O$_2$). In this case the magnetic expansion is not required and no restrictive conditions in pressure were observed.
 The DL formation appears to be related to the presence of negative ions and does not depend on specific atomic or molecular processes. The  potential drop across the DL  is independent of the electron temperature and is about 8 V. Unlike the electropositive DL, the electronegative DL is a loose structure the shape of which is extremely sensitive to either the experimental parameters (gas mixture, pressure...) or  intrusive diagnostics. Moreover, the electronegative DL becomes propagates at higher negative ion fractions. Finally, the electronegative DL could be traversed by the Langmuir probe without being destroyed which may be a consequence of its loose structure.

\section{Acknowledgments}

The authors would like to gratefully thank Prof. A.J. Lichtenberg for fruitful discussions and careful reading of the manuscript. Operation of the RFEA in the reactor was made possible
thanks to J. Miro-Padovani during his two months stay at LPTP. This work has been supported by the European Space Agency, under Ariadna
study contract ACT-04-3101.

\newpage

\end{document}